\documentclass[12pt]{article}
\usepackage{graphicx}
\usepackage{amssymb,amsmath}

\newcommand\pubnumber{ }
\newcommand\pubdate{\today}

\def\napoli{J.~Stefan Institute, Jamova 39, P.O.~Box 3000, 1001 Ljubljana, 
Slovenia}
\def\support{\footnote{This work is supported in part by the EU under contract MTRN-CT-2006-035482 (FLAVIAnet) and by the Slovenian Research Agency.}}

\def\Title#1{\begin{center} {\Large #1 } \end{center}}
\def\Author#1{\begin{center}{ \sc #1} \end{center}}
\def\Address#1{\begin{center}{ \it #1} \end{center}}

\newcommand\pubblock{\rightline{\begin{tabular}{l} \pubnumber\\
         \pubdate  \end{tabular}}}
\newenvironment{Abstract}{\begin{quotation}  }{\end{quotation}}
\newenvironment{Presented}{\begin{quotation} \begin{center} 
             PRESENTED AT\end{center}\bigskip 
      \begin{center}\begin{large}}{\end{large}\end{center} \end{quotation}}
\def\Acknowledgements{\bigskip  \bigskip \begin{center} \begin{large}
             \bf ACKNOWLEDGEMENTS \end{large}\end{center}}

\begin{document}
\begin{titlepage}
\pubblock

\vfill
\Title{Theory of $b \to s/d \, \nu \bar \nu$ }
\vfill
\Author{Jernej F. Kamenik\support}
\Address{\napoli}
\vfill
\begin{Abstract}
The $b \to s \,\nu\bar\nu$ transitions are sensitive probes of new physics (NP) in the form of non-standard Z penguin effects. They involve four experimentally accessible  observables, among which the
inclusive rate of $B \to X_s \nu\bar\nu$  is the most theoretically clean, but is challenging to study experimentally. Theory errors in exclusive rates are dominated by form factors' normalization. They can be reduced in rate ratios like the $K^*$ polarization fractions in $B\to K^*\nu\bar\nu$ or by studying the ratio $Br(B^-\to K^- \nu\bar\nu)/Br(B^- \to K^- \ell^+\ell^-)$. Measurable NP effects in $b \to s/d\, \nu\bar\nu$ can be expressed in terms of two real parameters  and are generally correlated with other flavor observables. Although in principle even in minimal flavor violating scenarios, NP can still saturate present direct bounds, in many explicit NP models new effects are already constrained by other B physics observables to be much smaller. Alternatively, $b \to s/d \,E_{\rm miss}$ can receive contributions from particles other than neutrinos in the final state and strong modifications of the invariant mass spectra are possible. However, interpretation of such effects in terms of bounds on interactions of such new particles is nontrivial 
when experimental searches employ kinematical cuts or their signal reconstruction efficiencies depend on expected signal kinematical distributions.
\end{Abstract}
\vfill
\begin{Presented}
CKM 2010 Conference\\
Warwick, England,  September 6--10, 2010
\end{Presented}
\vfill
\end{titlepage}
\def\thefootnote{\fnsymbol{footnote}}
\setcounter{footnote}{0}

\section{Introduction}

Rare B decays with a $\nu \bar\nu$ pair in the final state are among the theoretically cleanest flavor changing neutral current (FCNC) processes of B mesons. This is due to the absence of long distance QCD effects associated with photonic penguin amplitudes which dominate the low di-lepton invariant mass region in B decays with a charged lepton pair in the final state.

With the advent of Super-B facilities~\cite{Bona:2007qt,Aushev:2010bq}, the prospects of measuring the exclusive $B\to K\nu\bar\nu$ and $B\to K^*\nu\bar\nu$ branching ratios in the near future have become realistic. This has motivated a renewed theoretical interest~\cite{Altmannshofer:2009ma, Bartsch:2009qp} in these decays and in particular in their ability to test the short distance physics related to Z penguin amplitudes. 

In addition, the  experimental missing energy signature $B\to K^{(*)} E_{\rm miss}$ associated with the undetected neutrino pair also allows to probe physics beyond the standard model (SM) in the form of new light long-lived neutral particles, coupling to $\bar b s$ currents. Provided they are light enough to be produced in B decays associated with a $K^{(*)}$ and sufficiently long-lived to escape the detector, they will contribute to the $b \to s \nu \bar \nu$ observables, since the two final states cannot be distinguished experimentally.

\section{Effective Hamiltonian}

The effective Hamiltonian for $b \to s \nu \bar \nu$ transitions\footnote{The $b\to d\nu\bar\nu$ case can be recovered with the simple $s\to d$ quark flavor replacement.} is generally given by 
\begin{equation}
\mathcal H_{\rm eff} = -\frac{4G_F}{\sqrt 2} V_{tb} V^*_{ts} \left( C^\nu_L \mathcal O_L^\nu + C^\nu_R \mathcal O_R^\nu \right) + \rm h.c. \,,
\end{equation}
with the operators 
\begin{equation}
\mathcal O_L^\nu = \frac{e^2}{16\pi^2} (\bar s \gamma_\mu P_L b )(\bar \nu \gamma^\mu (1-\gamma_5)\nu)\, ,	\quad \mathcal O_R^\nu = \frac{e^2}{16\pi^2} (\bar s \gamma_\mu P_R b )(\bar \nu \gamma^\mu (1-\gamma_5)\nu) \,.
\end{equation}
In the SM,	$C_R^\nu$ is negligible while	$C_L^\nu	= -X (x_t )/ \sin^2 \theta_w ,$ where $x_t = m^2_t /m^2_W$. The most recent evaluation of the function $X(x_t)$~\cite{Brod:2010hi} at NLO in QCD and electroweak corrections yields $(C_L^\nu )_{\rm SM} = -6.33\pm 0.06$ where the error is dominated by the top mass uncertainty. 

\section{Observables}

\subsection{$B\to X_s \nu\bar\nu$}

The inclusive decay $B \to X_s \nu\bar\nu$ can be evaluated using heavy quark expansion and operator product expansion, and so offers the theoretically cleanest constraint on the Wilson coefficients $C_L^\nu$ and $C_R^\nu$. In the SM its dineutrino invariant mass ($\hat s\equiv(p_\nu + p_{\bar\nu})^2/m_b^2$) spectrum can be written as
\begin{equation}
\frac{d\Gamma(B\to X_s \nu\bar\nu)_{\rm SM}}{d\hat s} = m_b^5 \frac{\alpha^2 G_F^2}{128\pi^5} |V^*_{ts} V_{tb}|^2 \kappa(0) |(C_L^\nu)_{\rm SM}|^2 \mathcal S(m_s, \hat s)\,,
\label{eq:dGamma}
\end{equation}
where $\mathcal S(m_s, \hat s)$ and $\kappa(0)$ represent phase-space and virtual QCD corrections respectively. QCD corrections to the partonic rate are known at the one-loop order and yield $\kappa(0) = 0.83$~\cite{Grossman:1995gt, Buchalla:1995vs}. Leading $O(\Lambda^2/m_b^2)$ power corrections have also been determined~\cite{Grossman:1995gt, Falk:1995me} and result in an additional $3\%$ reduction of the integrated rate. Residual perturbative \& non-perturbative uncertainties have been estimated at 5\%~\cite{Altmannshofer:2009ma}.  A common approach to further reduce the parametric uncertainties related to the value of the b quark mass is to normalize eq. (\ref{eq:dGamma}) to the inclusive semileptonic decay rate $\Gamma(B \to X_c e \bar \nu)$. However, in this approach an additional uncertainty is introduced through the dependence of the semileptonic phase space factor on the charm quark mass. Recently a novel approach has been proposed~\cite{Altmannshofer:2009ma} which refrains completely from this normalization and employs eq. (\ref{eq:dGamma}) directly in combination with the $b$ quark mass in the 1S scheme~\cite{Hoang:1998ng
}, which is known to a precision of about 1\%~\cite{Bauer:2004ve}. Among the remaining parametric uncertainties in the SM, the CKM matrix elements $|V_{ts}^*V_{tb}|$ contribute another $3\%$ in the branching ratio (using~\cite{Bona:2009cj}). All together, this leads to a precise prediction $Br(\bar B^0 \to X_s \nu\bar\nu)_{\rm SM} = (2.7\pm0.2)\times 10^{-5}$~\cite{Altmannshofer:2009ma} \,. Unfortunately, the inclusive decay mode is very challenging to probe experimentally and the present bound from LEP~\cite{Barate:2000rc} will be difficult to improve in the near future. 

\subsection{$B\to K^{(*)} \nu\bar\nu$}

The exclusive $B^+\to K^{+} \nu\bar\nu$ mode presently exhibits the greatest sensitivity to $C_L^{\nu}$, bounding it to be within a factor of three compared to the SM value with $Br(B^+\to K^{+} \nu\bar\nu)<1.4\times 10^{-5}$~\cite{:2007zk}. A positive SM signal at a Super-B factory is expected at an integrated luminosity of $10$~ab${}^{-1}$ -- five times less than for the $B\to K^{*} \nu\bar\nu$ mode~\cite{Bona:2007qt}, which is presently bounded at $Br(B\to K^*\nu\bar\nu)<8.0\times 10^{-5}$~\cite{:2008fr}. On the other hand, the $K^*$ in the final state offers an additional independent observable in the form of its longitudinal or transverse polarization fraction~\cite{Altmannshofer:2009ma}
\begin{equation}
F_{L,T} = \frac{d\Gamma_{L,T}/d \hat s}{d\Gamma/ d\hat s}\,, \quad F_L = 1-F_T\,.
\end{equation}
Experimentally, these are accessible through the angular distributions of the $K^*$ decay products (e.g. $K\pi$)
\begin{equation}
\frac{d^2\Gamma}{d\hat s d\cos \theta} = \frac{3}{4} \frac{d\Gamma_T}{d\hat s} \sin^2\theta + \frac{3}{2} \frac{d\Gamma_L}{d\hat s} \cos^2\theta\,,
\end{equation}
where $\theta$ is the angle between the $K^*$ flight direction in the $B$ rest frame and the $K$ flight direction in the $K\pi$ rest frame.

The main theoretical uncertainty in the exclusive modes is due to the normalization and the shape of the relevant form factors. Presently the most precise calculations of these are based on QCD sum rule techniques~\cite{Ball:2004ye
} resulting in a sizable uncertainty for the branching ratios. Fortunately, these uncertainties partially cancel in the polarization fractions which can therefore be predicted with greater accuracy, i.e. the integrated longitudinally polarized fraction is predicted to be $\langle F_L\rangle_{\rm SM} = 0.54\pm 0.01$~\cite{Altmannshofer:2009ma}. 

In the case $B\to K \nu\bar\nu$ a combined analysis with $B \to K \ell^+\ell^-$ has recently been proposed~\cite{Bartsch:2009qp}. It allows to drastically reduce form factor uncertainties in the ratio of the two branching ratios once the narrow resonance region in $B \to K \ell^+\ell^-$ has been removed using appropriate cuts in di-lepton invariant mass.  Namely, although a new form factor ($f_T$) associated with the photonic penguin operator matrix element enters the theoretical prediction for the $B \to K \ell^+\ell^-$ rate, it can be related to the $f_+$ form factor determining $B\to K \nu\bar\nu$ in both the hard (when $E_K \simeq m_B/2$ in the B frame) and soft (when $|{\bf p}_K|\simeq0$) kaon limits at leading order in the heavy quark expansion and up to perturbative QCD corrections~\cite{Bartsch:2009qp, Kamenik:2009ze}. The remaining power corrections and so called weak annihilation contributions can be estimated using QCD factorization at small $\hat s$~\cite{Beneke:2001at}, while broad resonance contributions at high $\hat s$ have been estimated using a sum over few states. Consequently, one can define the ratio $R \equiv Br(B^-\to K^- \nu\bar\nu)/Br(B^- \to K^- \ell^+\ell^-)$ in which the form factor normalization uncertainty almost completely cancels, leading to a precise SM prediction
$R_{\rm SM} = 7.59^{+0.41}_{-0.48}$~\cite{Bartsch:2009qp},
where the stated uncertainty is dominated by an estimate of higher order perturbative QCD corrections.

Another potential source of uncertainty for the $B^+\to K^{(*)+} \nu\bar\nu$ modes comes from the leptonic decays $B^+\to \tau^+\nu$ with the tau subsequently decaying as $\tau^+\to K^{(*)+}\bar\nu$~\cite{Kamenik:2009kc}. Although such contributions are formally of the order $G_F^4$, such suppression is compensated by the narrow $\mathcal O(G_F^2)$ width of the tau lepton when it goes on-shell. Furthermore, such tau-pole contributions, when integrated over the unmeasured neutrino momenta, cover most of the total phase space available in  $B^+\to K^{(*)+} \nu\bar\nu$ and exhibit no resonant features that would distinguish them from the short distance contributions. In fact, such tree-level tau-mediated long distance (LD) contributions yield $98\%$ of the $B^+\to \pi^+\nu\bar\nu$ rate in the SM. And although they are Cabibbo suppressed for the $B^+\to K^{(*)+}\nu\bar\nu$ modes, they still result in $12\%(14\%)$ enhancements in the total rates. They also affect the inclusive $B^+\to X_s \nu\bar\nu$ mode. Fortunately, with enough data on $B\to \tau\nu$, the tau-pole contributions to $B^+\to K^{(*)+}\nu\bar\nu$ could be subtracted experimentally~\cite{Bartsch:2009qp} since approximately~\cite{Kamenik:2009kc}
\begin{equation}
Br(B^+\to K^{(*)+}\nu\bar\nu)_{\rm LD} \approx Br(B^+\to \tau^+\nu) \times Br(\tau^+\to K^{(*)+}\bar\nu)\,.
\end{equation}
Alternatively, they can also be computed and added to the theoretical prediction. This approach introduces parametric uncertainties due to the B meson decay constant and the $|V_{ub}|$ CKM matrix element entering the prediction for $Br(B^+\to \tau^+\nu)$\,. Presently, the resulting additional uncertainty on $Br(B^+\to K^{(*)+}\nu\bar\nu)$ is of the order $3\%(4\%)$ respectively~\cite{Kamenik:2009kc}. 

\section{New Physics Sensitivity}

The four observables accessible in the three different $b \to s\nu\bar\nu$ decays depend on the two in principle complex Wilson coefficients $C_L^\nu$ and $C_R^\nu$ . However, only two combinations of these complex quantities enter the relevant formulae and are thus observable. These are~\cite{Grossman:1995gt, Melikhov:1998ug}
\begin{equation}
\epsilon=\frac{\sqrt{|C_L^\nu|^2+ |C_R^\nu|^2}}{|(C_L^\nu)_{\rm SM}|}\,,\qquad \eta = \frac{-Re(C_L^\nu C_R^{\nu*})}{{|C_L^\nu|^2+ |C_R^\nu|^2}}\,.
\end{equation}
The experimental measurements of the branching ratios and $F_L$, can thus be translated to excluded areas in the $\epsilon-\eta$ plane, where the SM corresponds to $(\epsilon, \eta) = (1, 0)$. An important feature of $F_L$ in such an analysis is that it only depends on $\eta$~\cite{Altmannshofer:2009ma}, meaning that any deviation from the SM would imply the presence of right-handed currents.

\subsection{Minimal Flavor Violation}

In minimal flavor violating (MFV) scenarios~\cite{Chivukula:1987py}
, only $C^{\nu}_L$ can receive significant new contributions which are furthermore universal between $b \to s/d \, \nu\bar\nu$ modes. Furthermore, in models in which the bottom Yukawa effects can be neglected (i.e. models with a single Higgs boson or multi-Higgs models where the bottom Yukawa remains much smaller than the top), effects in $b \to s/d\, \nu\bar\nu$ transitions can be correlated with $s \to d\, \nu\bar\nu$ modes~\cite{Buras:2001af}. In particular, the existing measurement of $Br(K^+\to \pi^+\nu\bar\nu)$ already constrains $Br(B\to K^{(*)}\nu\bar\nu)$ to be within an order of magnitude of the SM values~\cite{Hurth:2008jc}. However even in these most constraining scenarios, present direct bounds on $Br(B\to K^{}\nu\bar\nu)$ are already stronger and thus represent a valuable constraint on such new physics (NP).

\subsection{Modified Z-penguins}

A complementary effective approach to NP in Z-penguin amplitudes considers possible modifications of the $\bar b s Z$ coupling~\cite{Buchalla:2000sk}. This leads to strong correlations with other rare $B$ decays, such as $B_s\to \mu^+\mu^-$ and $B\to X_s \ell^+\ell^-$. In particular the present measurements of $Br(B\to X_s \ell^+\ell^-)$ already constrain $b\to s \nu\bar\nu$ rates to be within a factor of two compared to SM predictions, unless several possible NP contributions to $B\to X_s \ell^+\ell^-$ conspire to reduce their combined effect in this mode~\cite{Altmannshofer:2009ma}. In a particular example~\cite{Buras:2010pz}, motivated by the resolution of the $S_{\psi\phi}$ puzzle, new right-handed sources of flavor violation are introduced, modifying the $\bar b s Z$ couplings accordingly. Then, the existing bounds from  $B\to X_s \ell^+\ell^-$ predict a strong anti-correlation between the $B\to K^{(*)}\nu\bar\nu$ rates -- only one of the two is expected to be enhanced.

\subsection{Minimal Supersymmetric Standard Model}

In the Minimal Supersymmetric Standard Model (MSSM) with a generic flavor violating soft sector there are various new contributions to the $b \to s \nu\bar\nu$ transition and one might expect that large effects are possible. However, once the existing constraints coming from other flavor changing processes are applied, the effects in $C_L^\nu$ and particularly in $C_R^\nu$ turn out to be quite limited in the MSSM. In particular, gluino contributions to both $C_{L,R}^\nu$ are highly constrained by the $b \to s\gamma$ decay and have only negligible impact~\cite{Grossman:1995gt, Yamada:2007me}. Similarly $\tan\beta$-enhanced Higgs contributions to $C_R^\nu$ are constrained by $B_s \to \mu^+\mu^-$~\cite{Altmannshofer:2009ma}. Finally, the largest contributions can be generated with up-squark - chargino loops with a single $(\delta_{u}^{RL})_{32}$ mass insertion~\cite{Buchalla:2000sk, Lunghi:1999uk}
. They only affect $C_L^\nu$ and can,  after taking into account existing bounds from $B\to X_s\gamma$, $B\to X_s\ell^+\ell^-$ and $\Delta m_s / \Delta  m_d$, enhance or suppress the $b \to s \nu\bar\nu$ rates by at most $35\%$~\cite{Altmannshofer:2009ma}. Larger deviations still seem to be possible if R-parity violating interactions are introduced~\cite{Kim:2009mp}.

\subsection{Light long-lived neutral particles}

Neutrinos are not detected in present experiments probing $b \to s/d \nu\bar\nu$ decays. Consequently various NP contributions with light neutral long-lived particles in the final state can mimic the experimental signature.  However, due to the modifications induced by such contributions, relations between the observables and the parameters $\epsilon$ and $\eta$ derived for pure $b \to s/d \nu\bar\nu$ decays will then no longer hold. The presence of such contributions would therefore be signaled by the failure of the individual constraints on the $\epsilon-\eta$ plane in meeting at a single point~\cite{Altmannshofer:2009ma}. Alternatively, if the new invisible particles have sizable masses, they would also manifest themselves through characteristic kinematical edges in the measured spectra. In turn, such spectral features need to be taken into account when interpreting existing experimental searches in terms of bounds on the effective operators coupling invisible particles to the $\bar b s$ currents~\cite{Smith}. Firstly because of kinematical cuts used to suppress backgrounds, but also to some extent because reconstruction efficiencies may depend on the final state kaon and/or pion momenta. Traditionally, experimental searches~\cite{:2007zk} have relied on SM predictions for the spectra to extract bounds on the $b \to s/d \nu\bar\nu$ decay branching ratios, but the  $B\to K^*\nu\bar\nu$ BaBar analysis~\cite{:2008fr} presents an example of how such model dependencies can be minimized\footnote{The author thanks Francesco Renga for instructive clarifications on this point.}.

\section{Conclusions}

The $b \to s \nu\bar\nu$ transitions are sensitive probes of NP in the form of non-standard Z penguin effects ($b \to d \,\nu\bar\nu$ decays of charged  $B$'s on the other hand are dominated by LD tau pole contributions). They involve four experimentally accessible  observables, among which the
inclusive rate of $B \to X_s \nu\bar\nu$  is the most theoretically clean, but is challenging to study experimentally.
Theory errors in exclusive rates are dominated by form factors' normalization uncertainties. They can be reduced in rate ratios like the $K^*$ polarization fractions in $B\to K^*\nu\bar\nu$ or by studying the ratio $ Br(B^-\to K^- \nu\bar\nu)/Br(B^- \to K^- \ell^+\ell^-)$.

Measurable NP effects in $b \to s/d \,\nu\bar\nu$ can be expressed in terms of two real parameters  and are generally correlated with other flavor observables. Although in principle even in MFV, NP can still saturate present direct bounds, in many explicit NP models new effects are already constrained by other B physics observables to be much smaller. Finally, $b \to s/d \,E_{\rm miss}$ can receive contributions from particles other than neutrinos in the final state and strong modifications of the invariant mass spectra are possible. However, interpretation of such effects in terms of bounds on interactions of such new particles is nontrivial when experimental searches employ kinematical cuts or their signal reconstruction efficiencies depend on expected signal kinematical distributions.
\vspace{-0.5cm}
\Acknowledgements
The author would like to thank the organizers for the invitation and hospitality at this very stimulating and interesting conference.

\end{document}